\documentclass[aip,reprint,pop]{revtex4-1}

\usepackage{graphicx}
\usepackage{subfigure}
\usepackage{epstopdf}
\usepackage{amsmath}
\usepackage{amssymb}
\usepackage{nicefrac}
\usepackage{dcolumn}
\usepackage{bm}
\usepackage[colorlinks=true, linkcolor=blue]{hyperref}

\begin{document}

\title{DC electric field generation and distribution in magnetized plasmas}
\author{Jean-Marcel Rax}
\affiliation{Andlinger Center for Energy + the Environment, Princeton
University, Princeton, NJ 08540, USA}
\affiliation{ IJCLab, Universit\'{e} de Paris-Saclay, 91405 Orsay, France}
\author{Renaud Gueroult}
\affiliation{LAPLACE, Universit\'{e} de Toulouse, CNRS, INPT, UPS, 31062 Toulouse, France}
\author{Nathaniel J. Fisch}
\affiliation{Department of Astrophysical Sciences, Princeton University, Princeton,NJ 08540, USA}

\begin{abstract}
Very large DC and AC electric fields cannot be sustained between conducting electrodes because of volume gas breakdown and/or surface field emission. However, very large potential fields are now routinely generated in plasma structures such as laser generated wake in unmagnetized plasmas. In magnetized plasmas, large DC fields can also be sustained and controlled perpendicular to the magnetic field, but the metallic end plates limiting the plasma, terminating the magnetic field lines and usually providing the voltage drop feed between the field lines, impose severe restrictions on the maximum field. However, it is shown that very large radial DC voltage drops can be sustained by injecting waves of predetermined frequencies and wave vectors, traveling along the azimuthal direction of an axially magnetized plasma cylinder, or by injecting fast neutral particles beams along this azimuthal direction. The large conductivity along the magnetic field lines and the small conductivity between the field lines then distribute this voltage drop. The global power balance and control parameters of wave and beam generated large DC electric fields in magnetized plasmas are identified, described and analyzed.
\end{abstract}

\date{\today}
\maketitle

\section{Introduction}

The quest for very large electric fields is mainly driven by the need for
more compact particles accelerators, but it is also important in other
fields such as: (\textit{i}) mass separation envisioned for nuclear waste
cleanup\cite{Gueroult2015}, spent nuclear fuel reprocessing~\cite{Dolgolenko2017,Timofeev2014,Gueroult2014a,Vorona2015,Yuferov2017,Litvak2003} and rare earth elements recycling~\cite{Gueroult2018a}, (\textit{ii}) advanced $E$ cross $B$ plasma
configurations for the purpose of ions acceleration~\cite{Janes1965,Janes1965a,Janes1966}, and (\textit{%
iii}) thermonuclear fusion with rotating tokamak~\cite{Rax2017,Ochs2017b} or rotating
mirrors~\cite{Hassam1997,Fetterman2010,Fetterman2008,Teodorescu2010,Bekhtenev1980}.

Two fields configurations can sustain a DC electric field in a magnetized
plasma : (\textit{i}) the \textit{Brillouin configuration} with an axial
magnetic field and a radial electric field and (\textit{ii}) the \textit{%
Hall configuration} with a radial magnetic field and an axial electric
field. This last configuration is the one at work in stationary plasmas
thrusters where ions are unmagnetized; the former one, where ions are
magnetized, is used in mass separator devices and advanced thermonuclear
traps.

This study is devoted to this last type of configuration. Brillouin type of
rotating plasmas have been widely studied since the early proposal of
Lehnert to take advantage of the isopotential character of magnetic field
lines and surfaces to sustain a voltage drop through external biasing at the
edge of a plasma column with concentric electrodes~\cite{Lehnert1970,Lehnert1973,Lehnert1974,Wilcox1959,Lehnert1971}. These rotating configurations have since then been explored both
theoretically and experimentally for mass separation~\cite{Rax2019,Kolmes2019,Krishnan1981,Ohkawa2002,Shinohara2007,Gueroult2014,Gueroult2016a,Zweben2018,Gueroult2019,Fetterman2009,Fetterman2011,Liziakin2020,Liziakin2021,Liziakin2022}, thermonuclear
confinement~\cite{Hassam1997,Fetterman2010,Fetterman2008,Teodorescu2010,Bekhtenev1980} and  the study astrophysical phenomena in laboratory experiments~\cite{Flanagan2020,Desangles2021}.

In this new study, rather than focusing specifically on separation or fusion
applications, we will address the generic issues of the power balance and
the field structure of unconventional radial electric field sustainment,
with waves or neutral beams, in a cylindrical plasma shell confined in a
magnetized column. We will present new promising results in terms of
efficiency and control of these advanced wave and beam schemes.

Three mains principles can be considered with respect to very high electric
field generation:
\begin{itemize}
\item[(\textit{i})]Accelerator technologies~\cite{Wiedemann2015} such as electrostatic, Van de
Graff type, accelerators where metallic electrodes are charged up to create a voltage drop of typically a few MV. These DC type of devices are limited by
electrons emission at metallic surfaces under high electric fields and/or
breakdown of the insulating gas. Modern RF and microwave accelerators bypass
this drawback of metallic surface through the use of high frequency fields
and can reach far higher AC electric fields values, but even at high
frequencies, metallic structures display an unavoidable electric field
threshold above which massive electrons emission takes place.

To address breakdown and emission problems, the use of fully ionized
plasma has been put forward.
\item[(\textit{ii})] Laser-Plasma accelerators bypass these problems through the
use of plasma rather than metals to sustain the electric charges separation,
and have reached voltage gradients in the GV per meter range. The basics of
such schemes is the generation of a travelling electrons-ions charge
separation with the ponderomotive force of an ultrashort laser pulse acting
on the electron population. Indeed, a short laser pulse of length $L$,
described by its vector potential $A$, will push the electrons in the
propagation direction and generate a charge separation with amplitude $%
q^{2}A^{2}L/2m^{2}c^{2}$~\cite{Rax1993,Rax1992}, where $q$ and $m$ are the electron
charge and mass and $c$ the velocity of light. Such a charge separation, of
the order of tens of $\mu $m in underdense plasmas, generates large
traveling fields which then will oscillate at the electron plasma frequency $%
\omega _{pe}$ behind the pulse as a wake. A well phased, and well shaped,
charged particles bunch, following the laser pulse, can gain energy in such
a laser generated electrostatic waves.
\item[(\textit{iii})] Besides these mature conventional and advanced accelerator
technologies, an overlooked physical principle can be put at work to
generate large DC electric field : using a magnetized plasma in which we
induce a steady state charge separation perpendicular to the magnetic field
through the continuous absorption of a resonant wave or the continuous
ionization of a fast neutral beam.
\end{itemize}

That a magnetic field can inhibit the relaxation of the charges separation
sustaining a very large voltage drop across a magnetic field is suggested by
the energy associated with both electric and magnetic fields : (\textit{i}) $%
\varepsilon _{0}E^{2}V/2$ for an electric field $E$ in a volume $V$ and (%
\textit{ii}) $B^{2}V/2\mu _{0}$ for a magnetic field $B$ in a volume $V$. A
large electric field of say $10$ [MV/m] is associated with a density of
energy (pressure) of the order of few [kJ/m$^{3}$], although a typical
magnetic field of say $1$ [T] is associated with a density of energy
(pressure) of the order of few [MJ/m$^{3}$]. This very strong ordering
between magnetic and electric pressure suggests why the free charges, which
are attached to the magnetic field through the cyclotron motion, can resist
the tendency to relaxation and (quasi-) neutralization driven by an electric
field perpendicular to the magnetic field.

\begin{figure*}
\begin{center}
\includegraphics[width=14cm]{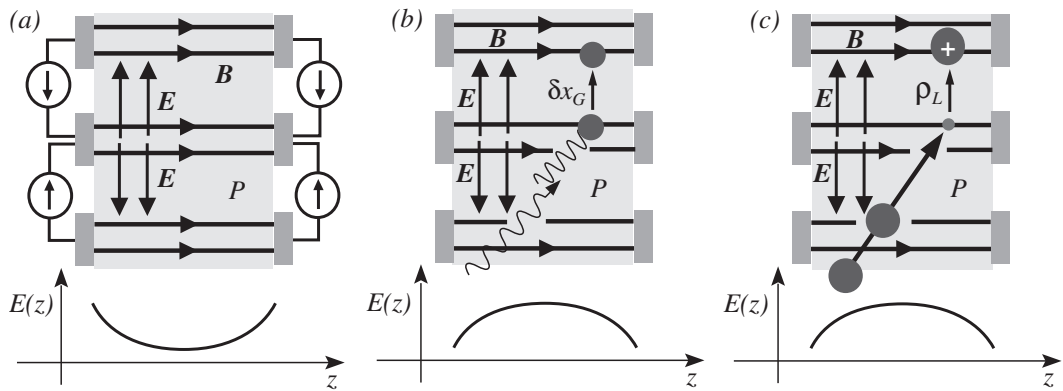}
\caption{(a) The classical method to sustain a perpendicular electric field
in a magnetized plasma (P) column with biased edge electrodes, (b) wave
driven charge separation  in a magnetized plasma (P) and (c) beam driven charge
separation  in a magnetized plasma (P). $E\left( z\right) $ is the radial
electric field between the axis and the outer cylindrical shell.}
\label{Fig:Fig1}
\end{center}
\end{figure*}

The wave and beam schemes considered in this study to drive an
electric field in a magnetized plasma are to be compared with the more classical scheme where a voltage drop between field lines is imposed with external voltage generators connected to the field
lines edges, as illustrated on Fig.~\ref{Fig:Fig1}(a). As we will demonstrate, an important conceptual difference is that in the classical scheme the electric field $E\left( z\right) $
has to penetrate the plasma column from the edge, and is decreasing along the 
$z$ axis from the left and right edges toward the center. On the other hand, wave or beam power can in principle be deposited at the center of a plasma column, as shown respectively in Figure~\ref{Fig:Fig1}(b) and Figure~\ref{Fig:Fig1}(c). In these new schemes the
maximum voltage drop thus occurs in the center while the minimum voltage drop is found the endplates, in contrast with the classical scheme. By allowing the electric field to be localized more inside the plasma than at the edge, with a weaker interaction with any solid material, the risk of breakdown and emission near metallic endplates are reduced, and larger values can be envisioned.

%Besides waves and beams considered here, the usual way to sustain an electric field in a magnetized plasma is simply to use the fact that the conductivity along field lines is very large and to impose a voltage drop between field lines with external voltage generators connected to the field lines edges as illustrated on Fig.~\ref{Fig:Fig1}(a). Figure~\ref{Fig:Fig1}(a) illustrates this usual way to sustain a DC electric field, this method takes advantage of the high conductivity along the field lines and voltage generators sustain a voltage drop between these conducting magnetic filed lines. Figure~\ref{Fig:Fig1}(b) illustrates the centrally wave driven method described and yes here. Figure~\ref{Fig:Fig1}(c) illustrates the centrally beam driven method described and analyzed here.

%As we will demonstrate, the main difference between (a) and (b,c) methods is the fact that in the classical one the electric field $E\left( z\right) $ has to penetrate the plasma column from the edge and is decreasing along the $z$ axis from the left and right edges toward the meridional zone; although, if wave or beam power is deposited at the center of a plasma column, the maximum voltage drop occurs in the center, not at the edge as previously, and the value of this voltage drop is minimum at the endplates as opposed to the classical methods. The fact that the electric field can be localized more inside the plasma than at the edge, with a weaker interaction with any solid material provides an efficient way to avoid breakdown and emission near metallic endplates, so that larger values can be envisioned

Practically, the upper limit for the amplitude of electric field generated
by a laser pulse in underdense plasmas is known to be associated with the occurrence of
cavitation behind the pulse. This phenomena has been observed numerically
and experimentally. On the other hand, the upper limit for the amplitude of the DC electric
field generated by wave or beam power absorption in magnetized plasmas has
never been explored. Moreover, the possibility to isolate this large DC
electric field from the plasma facing end plate in order to avoid breakdown
or electron emission has never been considered. Both of these issues are considered here. We will identify the constraint
arising from the plasma (\textit{i}) inherent anisotropic dissipation and (%
\textit{ii}) finite size, and then translate it into realistic conditions
for large field generation, distribution and dissipation, thus identifying
upper bounds on power consumption for DC high voltage generation across
magnetized plasmas. We will show that upper bounds in the GV/m range can be envisioned from
the proposed models of waves and beam generation under optimal conditions,
but that a few MV/m already provides the necessary conditions for the very fast supersonic
rotations of a fully ionized hot plasma columns (required for instance in thermonuclear
trap) and is accessible with wave or beam power of the order of few tens of
MW.

This paper is organized as follows. First, in section~\ref{Sec:SecII}, we present a
heuristic view of the formation of a voltage drop using waves and beams,
and address the issue of dissipation in a magnetized plasma. Then, in section~\ref{Sec:SecIII}, we briefly review the principle of charge transport driven by resonant
waves in a magnetized plasma, and identify from these results an upper bound for
DC electric field wave driven generation. Then, in section~\ref{Sec:SecIV}, we describe
the principle of charge separation driven by fast neutral beam injection.
The expression of the sustained DC electric field is established through
three different methods giving the very same result. The order of magnitude
of the maximum achievable electric field through this method is also
estimated. The steady state balance between wave/beam driven charge separation/generation and dissipative charge dispersion and (quasi-) neutralization is considered in section~\ref{Sec:SecV}. Specifically, a steady state model is obtained by considering the balance between (\textit{i}) wave/beam driven charge separation/generation, (\textit{ii}) fast distribution/spreading along the
field lines and (\textit{iii}) slow relaxation across the field lines. This model is then solved in section~\ref{Sec:SecVI} to identify both the plasma resistance $R$ and the
attenuation length $\lambda $ which describe the steady state of a wave, or
beam, driven magnetized and polarized plasma slab. The results are then used to address in section~\ref{Sec:SecVII} the issue of finite size plasmas in the case where the
attenuation length is too long to ensure a good confinement of the electric
field near the wave or beam active plasma zone and away from the plasma edges.
We show that a decrease the voltage drop at the edge of the plasma can be
achieved at the cost of a certain loss of the efficiency of the generating
process. Finally, the last section, section~\ref{Sec:SecVIII}, summarizes our new findings and point towards the optimization of these DC electric field generation and confinement schemes
when additional constraints are considered, either for thermonuclear control
in rotating mirrors or mass separation purposes.

\section{Formation of voltage drop inside a magnetized plasma}

\label{Sec:SecII}

This section provides a heuristic presentation of the problem of electric
field generation in a plasma.

\begin{figure}
\begin{center}
\includegraphics[width=5cm]{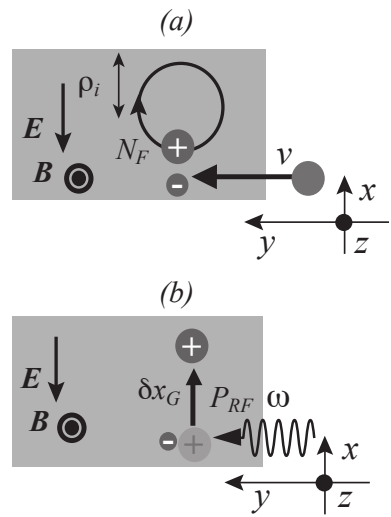}
\caption{(a) Neutral beam driven perpendicular electric polarization and (b)
wave driven perpendicular electric current generation.}
\label{Fig:Fig2}
\end{center}
\end{figure}

Consider a magnetized plasma and a Cartesian set of coordinates $\left(
x,y,z\right) $ and a Cartesian basis $\left( \mathbf{e}_{x},\mathbf{e}_{y},%
\mathbf{e}_{z}\right) $. A wave propagating along the $y$ direction,
perpendicular to the magnetic field $B\mathbf{e}_{z}$, with wave vector $%
k_{\perp }\mathbf{e}_{y}$ and frequency $\omega $, generates a charge
separation of the resonant population and pushes each resonant particle by
an amount

\begin{equation}
\delta x_{G}=\frac{k_{\perp }}{q\omega B}\delta \mathcal{E}  \label{deltaxg}
\end{equation}
where $\delta \mathcal{E}$ is the amount of energy absorbed by the resonant
particle and $x_{G}$ its guiding center position. This process is
illustrated on Fig.~\ref{Fig:Fig2}(a).

When the quantum of energy $\delta \mathcal{E}$ = $\hbar \omega $ is
absorbed, the quantum of perpendicular momentum $\hbar k_{\perp }$ along $y$
is also absorbed and through a continuous absorption this provides a secular
force $\hbar k_{\perp }/\delta t$ which drives a drift along $x$ : $\hbar
k_{\perp }/\delta tqB$. During a time $\delta t$ the shift in
position is thus equal to $\hbar k_{\perp }/qB$, which eliminating $\hbar=\delta \mathcal{E}$/$\omega $ gives Eq.~(\ref{deltaxg}). This
relation Eq.~(\ref{deltaxg}) will be reviewed in the next section.

If, rather than $\delta \mathcal{E}\left[ \text{J}\right] $, we consider a
stationary (density of) power absorption $P_{RF}\left[ \text{W/m}^{3}\right] 
$, then Eq.~(\ref{deltaxg}) shows that a continuous wave drive will generate
a continuous guiding center current density $J_{\perp }\mathbf{e}_{x}$
perpendicular to the magnetic field

\begin{equation}
J_{\perp }\left[ \frac{\text{A}}{\text{m}^{2}}\right] =\frac{k_{\perp }}{%
\omega B}\cdot P_{RF}\left[ \frac{\text{W}}{\text{m}^{3}}\right]
\label{current}
\end{equation}
where $P_{RF}$ is the density of power absorbed by the resonant population.
This perpendicular drift current generation has been proposed to confine
toro\"{i}dal plasmas~\cite{Rax2017,Ochs2017b} and, for unstable waves, to provide a free
energy extraction mechanism from thermonuclear plasmas through \textit{alpha
channeling} both in tokamaks and mirrors~\cite{Fetterman2008,Fisch1992,Fisch1994,Fisch1995,Herrmann1997}.

Rather than a wave, we consider now a fast neutral beam as a momentum
source, with velocity $v\mathbf{e}_{y}$, injected in a magnetized plasma as
illustrated in Fig.~\ref{Fig:Fig2}(b). When a fast neutral particle is ionized inside
the plasma, the electron and the ion rotate in opposite direction and the
value of their Larmor radius is so different that these two charges are
separated on average by an amount 
\begin{equation}
\delta x_{G}\approx \frac{Mv}{qB}=\rho _{i}\gg \rho _{e}  \label{polz2}
\end{equation}
where $\rho _{e/i}$ is the electron/ion Larmor radius and $M$ and $q$ are the
ion mass and charge.

The balance between the ionization rate of the fast neutral and the slowing
down of the fast ions provides a steady state density of fast ions $N_{F}$.
The associated steady state charge separation can be described by an
electric polarization $P_{\perp }\mathbf{e}_{x}$ perpendicular to the
magnetic field
\begin{equation}
P_{\perp }\left[ \frac{\text{C}}{\text{m}^{2}}\right] =\frac{Mv}{B}\cdot
N_{F}\left[ \frac{\text{1}}{\text{m}^{3}}\right]  \label{polz}
\end{equation}
This electric polarization $P_{\perp }$ is the source of a voltage drop
between magnetic filed lines, which will be analyzed in section~\ref{Sec:SecIV}.

In this study we will identify, describe and analyze schemes to use this
wave driven current $J_{\perp }$ Eq.~(\ref{current}) or this beam driven
polarization $P_{\perp }$ Eq.~(\ref{polz}) to generate a large voltage drop
across the magnetic field lines in the core of the plasma. Core generation
provides a way to mitigate breakdown and/or emission at the edge of the
plasma when both the plasma and the field lines encounter the end plates.

A picture of the build-up phase of a growing electric field in a plasma slab can be
described as follows. Note that in the following model we do not consider the
interplay between the adiabatic and resonant response of the particles~\cite{Ochs2021a,Ochs2022,Ochs2021}, and consider the final global momentum balance. A wave, or a neutral beam, moves some minority charges across the magnetic field as shown by Eqs.~(\ref{deltaxg}, \ref{polz2}), and thus sets up a current $\mathbf{J}_{0}\left(
t\right) $ such that $\mathbf{J}_{0}\left( t=-\infty \right) $ = $\mathbf{0}$ 
 and $\mathbf{J}_{0}\left( t=0\right) $ = $\mathbf{J}_{0}$ (dissipation is
switched off for $t<0$). From an electrical point of view this phase
correspond to a capacitive electric field build up in a non dissipative
dielectric media : the charging of a capacitor. The plasma, which displays a
low frequency permittivity $\varepsilon $ = $1+\omega _{pi}^{2}/\omega
_{ci}^{2}$ $\approx \omega _{pi}^{2}/\omega _{ci}^{2}$, adjusts an electric
field $\mathbf{E}\left( t\right) $ such that the electrostatic limit of
Maxwell-Amp\`{e}re equation is fulfilled 
\begin{equation}
\varepsilon _{0}\frac{\omega _{pi}^{2}}{\omega _{ci}^{2}}\frac{\partial 
\mathbf{E}}{\partial t}+\mathbf{J}_{0}\left( t\right) =\mathbf{0}.
\end{equation}
From a mechanical point of view this build-up phase corresponds to a
momentum input through the $\mathbf{J}_{0}\left( t\right) \times \mathbf{B}$
force and this momentum ends up in the plasma $E$ cross $B$ drift, guaranteeing
momentum conservation 
\begin{equation}
\int_{-\infty }^{0}\mathbf{J}_{0}\left( t\right) \times \mathbf{B}dt+N_{p}M%
\frac{\mathbf{E}_{0}\times \mathbf{B}}{B^{2}}=\mathbf{0}
\end{equation}
where $\mathbf{E}\left( t=0\right) =\mathbf{E}_{0}$, $M$ is the ion mass and 
$N_{p}$ the ion density.

Then, for $t>0$ that is in the steady state dissipative regime, the charge
separation associated with $\mathbf{J}_{0}$ is short circuited by the plasma
conductivity through the conduction current $\mathbf{J}_{\text{conduction}}$
in the magnetized plasma, as well as the boundary condition at the edge of
the magnetic field lines. After this build up phase, the steady state is
reached when 
\begin{equation}
\mathbf{\nabla }\cdot \left( \mathbf{J}_{0}+\mathbf{J}_{\text{conduction}%
}\right) =0
\end{equation}

This steady state regime will be described within a framework where the
plasma is modeled as a slab of an anisotropic conductor, and the end
plates at the outer edges of the magnetic field lines will be modeled by a
resistive load $R_{L}$.

\begin{figure}
\begin{center}
\includegraphics[width=8cm]{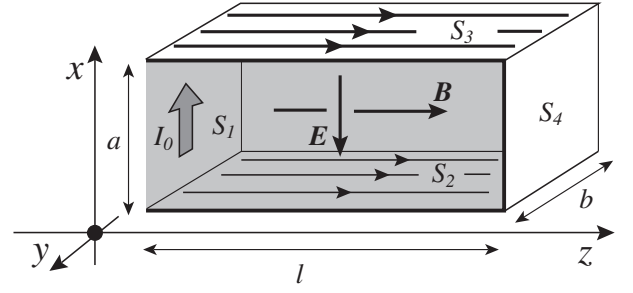}
\caption{A magnetized plasma slab $\left( a,b,l\right) $ with wave or beam
current drive $I_{0}$ localized on the left side $S_{1}$.}
\label{Fig:Fig3}
\end{center}
\end{figure}

Consider the magnetized plasma slab, illustrated on Fig.~\ref{Fig:Fig3}, with the
following dimensions : $a$ along $x$, $b$ along $y$ and $l$ along $z$. This
plasma slab is magnetized along $z$, $\mathbf{B}=B\mathbf{e}_{z}$, and we
assume that a wave or beam driven steady state electric current $I_{0}$ flows
along the face $S_{1}$ from the lower magnetic surface $S_{2}$ up to the
upper magnetic surface $S_{3}$. The two magnetic surface $S_{2}$ and $S_{3}$
are thus charged like a capacitor, but the electric conductivity along the
magnetic field line $\eta _{\shortparallel }$ and across the magnetic field
line $\eta _{\perp }\ll \eta _{\shortparallel }$ complexifies this simple
capacitor charging model and relaxes the stored charges. This conductive
charge redistribution and relaxation is the source of the voltage
distribution and power dissipation involved in the process of wave or beam
DC electric sustainment in a plasma identified and analyzed here.

The voltage drop along $S_{1}$ between $S_{2}$ and $S_{3}$ is $V_{0}$ so
that the power needed to sustain the steady state electric field ($V_{0}/a$) 
$\mathbf{e}_{x}$ near $S_{1}$ is simply $I_{0}V_{0}$. Two asymptotic cases
can to be considered in order to set up an equivalent circuit model.

First, if $S_{4}$ is a conductive short circuit between $S_{2}$ and $S_{3}$
the power $\mathcal{P}$ needed to sustain the steady state will be
approximately
\begin{equation}
\mathcal{P}_{\text{short cicuit}}=I_{0}V_{0}\approx \frac{l}{ab\eta
_{\shortparallel }}I_{0}^{2}=\frac{ab}{l}\eta _{\shortparallel }V_{0}^{2}
\end{equation}
as it is the conductivity along the magnetic field which will ensure
preferentially the charge relaxation at $S_{4}$. Second, if $S_{4}$ is non conductive, $S_{2}$ and $S_{3}$ are isolated and
the power needed to sustain the steady state will be approximately
\begin{equation}
\mathcal{P}_{\text{open cicuit}}=I_{0}V_{0}\approx \frac{a}{bl\eta _{\perp }}%
I_{0}^{2}=\frac{bl}{a}\eta _{\perp }V_{0}^{2}
\end{equation}
as the charge relaxation takes place across the magnetic field in the plasma
volume rather than at the edge. 

For a given voltage requirement $V_{0}$, and because $%
\eta _{\perp }$ $\ll $ $\eta _{\shortparallel }$, $\mathcal{P}_{\text{open
cicuit}}$ $\ll $ $\mathcal{P}_{\text{short cicuit}}$. In between these two
asymptotic limits, we will calculate the equivalent resistance of the slab $%
R_{e}$, Eq.~(\ref{equi}), and the power balance of the wave or beam
generation process Eq.~(\ref{pow678}). These are the main new results
presented in this article. The new expression for $R_{e}$ involves both what
we call the plasma resistance $R$ and a penetration length $\lambda $
describing the spatial decay of the voltage drop away from the source region.

\section{Wave-driven resonant charge separation}
\label{Sec:SecIII}

In this section we derive the relations Eqs.~(\ref{deltaxg}) and (\ref{current}) and briefly review the main relations describing the dynamics of wave
driven resonant charges separation in a plasma. This phenomena has been
proposed to provide free energy extraction in thermonuclear plasma~\cite{Fisch1992,Fisch1994,Fisch1995,Herrmann1997}
and to help toro\"{i}dal confinement in tokamak~\cite{Rax2017,Ochs2017b}.

The Cartesian plasma slab considered in the following is magnetized
along $z$, $\mathbf{B}=B\mathbf{e}_{z}$ and polarized along $x$, $\mathbf{E}%
=-E\mathbf{e}_{x}$. A wave with wave vector $\mathbf{k}$ = $k_{\perp }%
\mathbf{e}_{y}$ $+$ $k_{\parallel }\mathbf{e}_{z}$ and frequency $\omega $
propagates in this plasma along $\left( z\right) $ and across $\left(
y\right) $ the magnetic field. We restrict the following argument to an
unspecified components of this wave oscillating with the phase ($\omega t$ $%
- $ $k_{\perp }y$ $-$ $k_{\parallel }z$). In order to identify the
wave-particle resonances, we plug into the phase of this wave the
unperturbed motion of a charged particle characterized by the invariants ($%
x_{G}$, $v_{\parallel }$, $v_{c}$) 
\begin{eqnarray}
x &=&x_{G}+\frac{v_{c}}{\omega _{c}}\cos \left( \omega _{c}t\right) \text{,}
\\
y &=&\frac{E}{B}t+\frac{v_{c}}{\omega _{c}}\sin \left( \omega _{c}t\right) 
\text{,} \\
z &=&v_{\parallel }t.
\end{eqnarray}
Here $\omega _{c}$ is the cyclotron frequency, $v_{c}$ the cyclotron
velocity, $v_{\parallel }$ the velocity along the field lines and $x_{G}$
the guiding center position along $x$. The phase seen by a particle is thus
\begin{multline}
\cos \left( \omega t-k_{\perp }y-k_{\parallel }z\right) \sim \cos \left(
\omega t-k_{\perp }\frac{E}{B}t\right.\\
\left.-k_{\perp }\frac{v_{c}}{\omega _{c}}\sin
\omega _{c}t-k_{\parallel }v_{\parallel }t\right).
\end{multline}
This result can be rearranged with the classical Euler Bessel expansion 
\begin{equation}
\cos (a+b\sin \phi )=\sum_{N=-\infty }^{N=+\infty }\text{J}_{N}(b)\sin
(a+N\phi )
\end{equation}
so that the field seen by the particle becomes a series of harmonics with
Bessel function amplitudes 
\begin{multline}
\cos \left( \omega t-k_{\perp }y-k_{\parallel }z\right) \sim \sum_{N=-\infty
}^{N=+\infty }\text{J}_{N}\left( k_{\perp }\frac{v_{c}}{\omega _{c}}\right)\\
\times
\sin \left( \omega t-k_{\perp }\frac{E}{B}t-N\omega _{c}t-k_{\parallel
}v_{\parallel }t\right).
\end{multline}
Thus a resonance might occur with the $N$ component of this spectral
expansion if this oscillating amplitude becomes stationary :
\begin{equation}
\omega -k_{\perp }E/B-N\omega _{c}-k_{\parallel }v_{\parallel }=0.
\label{res}
\end{equation}

When this condition is fulfilled the topology of the particles motion phase
portrait changes and particles trapped in the wave experience a large
variation of the invariants of the free motion $\left( x_{G},v_{\parallel
},v_{c}\right) $. When this condition is not fulfilled the particles
oscillate and this oscillation is associated with a reactive power so that
no active power is exchanged with non resonant (adiabatic) particles.

For such resonances, if an amount $\delta \mathcal{E}$ of RF energy is
absorbed by a resonant particle, then the unperturbed motion invariants $%
\left( x_{G},v_{\parallel },v_{c}\right) $ are no longer invariant. Because
of the resonant interaction with the wave they become $\left( x_{G}+\delta
x_{G},v_{\parallel }+\delta v_{\parallel },v_{c}+\delta v_{c}\right) $ where 
$\left( \delta x_{G},\delta v_{\parallel },\delta v_{c}\right) $ are
proportional to $\delta \mathcal{E}$, a simple dynamical analysis allows to
write the set of relations : 
\begin{eqnarray}
\delta x_{G} &=&\frac{k_{\perp }}{q\omega B}\delta \mathcal{E}\text{,}
\label{dfg} \\
m\delta v_{\parallel } &=&\frac{k_{\parallel }}{\omega }\delta \mathcal{E}%
\text{,}  \label{dfg22} \\
mv_{c}\delta v_{c} &=&N\frac{\omega _{c}}{\omega }\delta \mathcal{E}\text{.}
\label{cyc}
\end{eqnarray}
Equation (\ref{dfg}) is associated with the conservation of the canonical
momentum along $y$. Eq.~(\ref{dfg22})  is associated with the conservation of classical
momentum along $z$. Finally, Eq.~(\ref{cyc}) describes harmonic cyclotron heating.
These relations can be rederived from an Hamiltonian analysis~\cite{Rax2018}, or
simply from the quantum photon picture described in the previous section.

Global (wave + particle) energy conservation can be simply checked as
follows. The complete variation of a resonant particle kinetic $%
mv_{\parallel }\delta v_{\parallel }+mv_{c}\delta v_{c}$ and potential $%
qE\delta x_{G}$ energy is 
\begin{align}
qE\delta x_{G}+mv_{\parallel }\delta v_{\parallel }+mv_{c}\delta v_{c} & =\frac{%
\delta \mathcal{E}}{\omega }\left( \frac{k_{\perp }E}{B}+k_{\parallel
}v_{\parallel }+N\omega _{c}\right)\nonumber\\
 & =\delta \mathcal{E}
\end{align}
where we have used the resonance condition Eq.~(\ref{res}) to obtain the final identity.

From these results we can identify a theoretical maximum electric field $%
E^{*}$ that can be sustained \textit{in situ} in a plasma with this type of
resonant charge separation process. The optimal wave, such that all the
energy $\delta \mathcal{E}$ goes to the charge separation and ends up in the
form of potential, $qE\delta x_{G}$, rather than kinetic, $mv_{\parallel
}\delta v_{\parallel }+mv_{c}\delta v_{c}$, energy, is a wave displaying no
Landau and cyclotron absorptions such that $k_{\parallel }$ = $N$ = 0 (we do
not consider here anomalous Doppler resonances where the wave transfer
energy between degrees of freedom). Equation (\ref{res}) thus becomes a
simple drift resonance : $\omega $ $=$ $k_{\perp }E^{*}/B$. This last
relation is confirmed by the energy balance restricted to potential energy $%
\delta \mathcal{E}$ $=$ $qE^{*}\delta x_{G}$. Tthen, with the help of Eq.~(%
\ref{dfg}) we eliminate $\delta \mathcal{E}$ to find the constraint on the
DC electric field $E_{RF}^{*}$ :
\begin{equation}
\frac{E_{RF}^{*}}{B}=\frac{\omega }{k_{\perp }}.  \label{estar}
\end{equation}

Very large $E_{RF}^{*}$ can thus in principle be reached for very large $B$ field values, though it is to be noted that the wave dispersion $\omega \left( k_{\perp
}\right) $ is also a function of $B$. Taking a moderate value of $B$ of the
order of few tesla and a high frequency wave with a velocity of the order of
the velocity of light, which is the case in tenuous plasmas, we end up with
electric fields values of the order of 1GV/m. The relation Eq.~(\ref{estar})
however only offers a partial view of the problem because if we want to drive the plasma
drift motion we need waves with a large momentum $k_{\perp }$, whereas Eq.~(\ref
{estar}) suggest that small $k_{\perp }$ are preferable for large electric
field. Equation (\ref{estar}) is an upper bound associated with an optimal
use of the wave power in term of efficiency. It is a kinematical constraint
associated with optimal resonance. This large value is only achieved if
dissipation (charges relaxation) is neglected. In the following we will
assume that the wave driven charge separation takes place in a narrow region
around $z=0$ and that this RF region is hot and collisionless but the
neighboring region are assumed collisional, and we will analyze the impact of
dissipative charge relaxation in a plasma slab.

\section{Neutral-beam-driven charge separation}
\label{Sec:SecIV}

In this section we derive the relations Eqs.~(\ref{polz2}) and (\ref{polz})
and set up and solve a simple model describing beam driven charges
separation and electric field generation in a magnetized plasma. This
phenomena is illustrated on Fig.~\ref{Fig:Fig2}(a) : a beam of fast neutral atoms with
velocity $v\mathbf{e}_{y}$ and density $N_{B}$ is directed toward a plasma
magnetized with $\mathbf{B}=B\mathbf{e}_{z}$. These fast atoms are ionized
through collisions with the plasma electrons and ions and also through
charges exchange with slow ions. Both processes provide fast ions generation
from these fast neutral.

The rate of fast ion generation from fast neutral is $\nu $ and it takes
into account both ionization and charge exchange. As soon as a fast ion is
generated in the plasma, it start to slow down with a typical slowing down
time $\tau $. If we consider fast hydrogen atom in a thermonuclear $pB11$
plasma, $\tau $ also accounts for fast proton pitch angle scattering on boron
ions. The density of fast ions in the plasma, $N_{F}$, is thus given by the
solution of the particles balance

\begin{equation}
\frac{dN_{F}}{dt}=\nu N_{B}-\frac{N_{F}}{\tau }
\end{equation}
Considering a steady state injection, the relation between the density of
fast ions, i.e. ions with a large Larmor radius, and the density of injected
neutral is 
\begin{equation}
N_{F}=N_{B}\nu \tau  \label{part}
\end{equation}

Three methods are considered below to calculate the DC electric field
sustained by steady state neutral beam injection.

First, the conservation of linear momentum in the $y$ direction can be used to
calculate the electric field $E\mathbf{e}_{x}$ generated by the beam. If we
neglect the electron mass $m$ in front of the ion mass $M$, the beam density
of momentum $N_{B}Mv$ which is coupled to the plasma at a rate $\nu $
provide a density of force $N_{B}Mv\nu $. This density of force acts during
a time $\tau $ on the plasma. The corresponding density of momentum $%
N_{B}Mv\nu \tau \mathbf{e}_{y}$ is absorbed in the form of plasma linear momentum along $y$. If we write $N_{P}$ the plasma density the
linear momentum balance can be written : 
\begin{equation}
N_{B}Mv\nu \tau \mathbf{e}_{y}=N_{p}M\frac{E\mathbf{e}_{x}\times B\mathbf{e}%
_{z}}{B^{2}}  \label{mome}
\end{equation}

The very same relation can be obtained from an electrical analysis rather
than from a mechanical point of view. If we neglect the electron Larmor
radius in front of the ion Larmor radius, the steady state
density of fast ions $N_{F}$ is associated with an electric polarization Eq.
(\ref{polz}) $N_{F}q\rho _{i}\mathbf{e}_{x}$ = $N_{F}\left( Mv/B\right) 
\mathbf{e}_{x}$. In response to this electric polarization, the plasma, which
displays a low frequency permittivity $\varepsilon $ = $1+\omega
_{pi}^{2}/\omega _{ci}^{2}$ $\approx \omega _{pi}^{2}/\omega _{ci}^{2}$, sets
up a reverse polarization through an electric field generation $E\mathbf{e}%
_{x}$. The condition for this dielectric dipole screening is 
\begin{equation}
N_{F}\frac{Mv}{B}\mathbf{e}_{x}+\varepsilon _{0}\frac{\omega _{pi}^{2}}{%
\omega _{ci}^{2}}E\mathbf{e}_{x}=\mathbf{0}.
\end{equation}
Here $\omega _{pi}$ is the ion plasma frequency and $\omega _{ci}$ the ion
cyclotron frequency. Taking the cross product of this last relation with $%
\mathbf{B}$ we find the condition
\begin{equation}
-N_{B}\nu \tau Mv\mathbf{e}_{y}+MN_{p}\frac{E\mathbf{e}_{x}\times B\mathbf{e}%
_{z}}{B^{2}}=\mathbf{0},  \label{mom2}
\end{equation}
which is Eq.~(\ref{mome}).

Finally, as a third demonstration of this result, we can consider
Maxwell-Amp\`{e}re equation with (\textit{i}) the polarization current $d%
\mathbf{P}_{\perp }/dt$ = $\left( N_{B}Mv/B\right) \nu \mathbf{e}_{x}$,
describing the generation of fast ions and (\textit{ii}) the displacement
current $\varepsilon _{0}\varepsilon \partial \mathbf{E}/\partial t$ = $%
\varepsilon _{0}\varepsilon \mathbf{E}/\tau $ associated with the decay of
the electric field due to these fast ions slowing down. In writing
Maxwell-Amp\`{e}re equation we neglect the diamagnetic effect of the fast
ions and consider $\mathbf{B}_{\text{\textit{fast ions}}}$ = $\mathbf{0}$
such that $\mathbf{\nabla }\times \mathbf{B}_{\text{\textit{fast ions}}}$ = $%
\mathbf{0}$ which implies $\partial \mathbf{P}_{\perp }/\partial t$ + $%
\varepsilon _{0}\varepsilon \partial \mathbf{E}/\partial t$ = $\mathbf{0}$. In this case 
\begin{equation}
N_{B}\frac{Mv}{B}\nu \mathbf{e}_{x}+\varepsilon _{0}\frac{\omega _{pi}^{2}}{%
\omega _{ci}^{2}}\frac{E}{\tau }\mathbf{e}_{x}=\mathbf{0}  \label{mom3},
\end{equation}
which is again identical to Eqs.~(\ref{mome}) and (\ref{mom2}).

Thus, no matter the point of view, (\textit{i}) mechanical with the momentum
balance Eq.~(\ref{mome}), (\textit{ii}) electrostatic with the dielectric
dipole screening Eq.~(\ref{mom2}), and (\textit{iii}) electrodynamic with
Maxwell-Amp\`{e}re Eq.~(\ref{mom3}), we find that the continuous injection
of a neutral beam along $y$ will sustain a DC electric field along $x$ : 
\begin{equation}
\frac{E_{NB}}{B}=v\frac{N_{B}}{N_{p}}\nu \tau.  \label{ebeam 2}
\end{equation}

To obtain an order of magnitude estimate we can take values typical of large tokamak
plasmas experiments : $N_{B}/N_{p}$ $\sim $ $10^{-4}-10^{-5}$, $\nu \tau $ $%
\sim $ $10^{5}-10^{6}$ and $v\sim 10^{6}-10^{7}$ [m/s]. In all these
relations both $\nu $ and $\tau $ are average as they are function of the
neutrals and fast ions velocities. With these values, an upper bound of tens or up to a few
hundreds of MV/m is found for the DC electric
field generation in magnetized plasma with neutral beam. The power flux in
the plasma from the neutral beam is given by : $P_{NB}\left[ \text{W/m}%
^{2}\right] $ = $N_{B}Mv^{3}/2$ so that the electric field Eq.~(\ref{ebeam 2}%
) can be rewritten as 
\begin{equation}
E_{NB}\left[ \frac{\text{V}}{\text{m}}\right] =\frac{2B\nu \tau }{Mv^{2}N_{p}%
}\cdot P_{NB}\left[ \frac{\text{W}}{\text{m}^{2}}\right].  \label{ebeam}
\end{equation}
To identify the limit of this generation process we can consider the simple density requirements for the previous ionization/slowing down model, that is $%
N_{p}\geq 10\times N_{B}$. For this density ratio the maximum electric field achievable with this scheme is
\begin{equation}
\frac{E_{NB}^{*}}{B}=v\frac{\nu \tau }{10}.  \label{max5}
\end{equation}

Both relations Eq.~(\ref{estar}) and Eq.~(\ref{max5}) are ultimate upper
bound when the longitudinal and transverse conductivities of the finite size
plasma slab can be ignored and the power deposition is optimized. The
relations Eq.~(\ref{estar}) and Eq.~(\ref{max5}) provide rough estimates of
the theoretical maximum values achievable with waves and beams, and are not
associated with a breakdown threshold but with an optimal power deposition
processes. Importantly, these relations predict very large upper bounds for the electric field both for wave and beam driven schemes, typically larger than tens of MV/m. 

Because the typical values we have in mind for advanced high energy supersonic rotating
plasmas applications are in the range of few tens of MV/m, we can consider
the full picture for such configurations and address the issue of voltage
distribution in the next section. The issue of dissipation in the bulk of a
finite size plasma slab, far from the wave or beam active regions, is also
addressed in this coming section. Note finally that Eq.~(\ref{estar}) does not
involve dissipative time scales, whereas Eq.~(\ref{max5}) involves the
dissipative time scales $\nu $ and $\tau $. This difference is due to the
fact that a wave can kick thermal particles so that, if we ignore
temperature gradients, this does not perturb the thermal equilibrium. On the other hand, the fast ions must ultimately thermalize and isotropize in the neutral beam case.

\section{Voltage drop distribution in a plasma}
\label{Sec:SecV}

\begin{figure}
\begin{center}
\includegraphics[width=8cm]{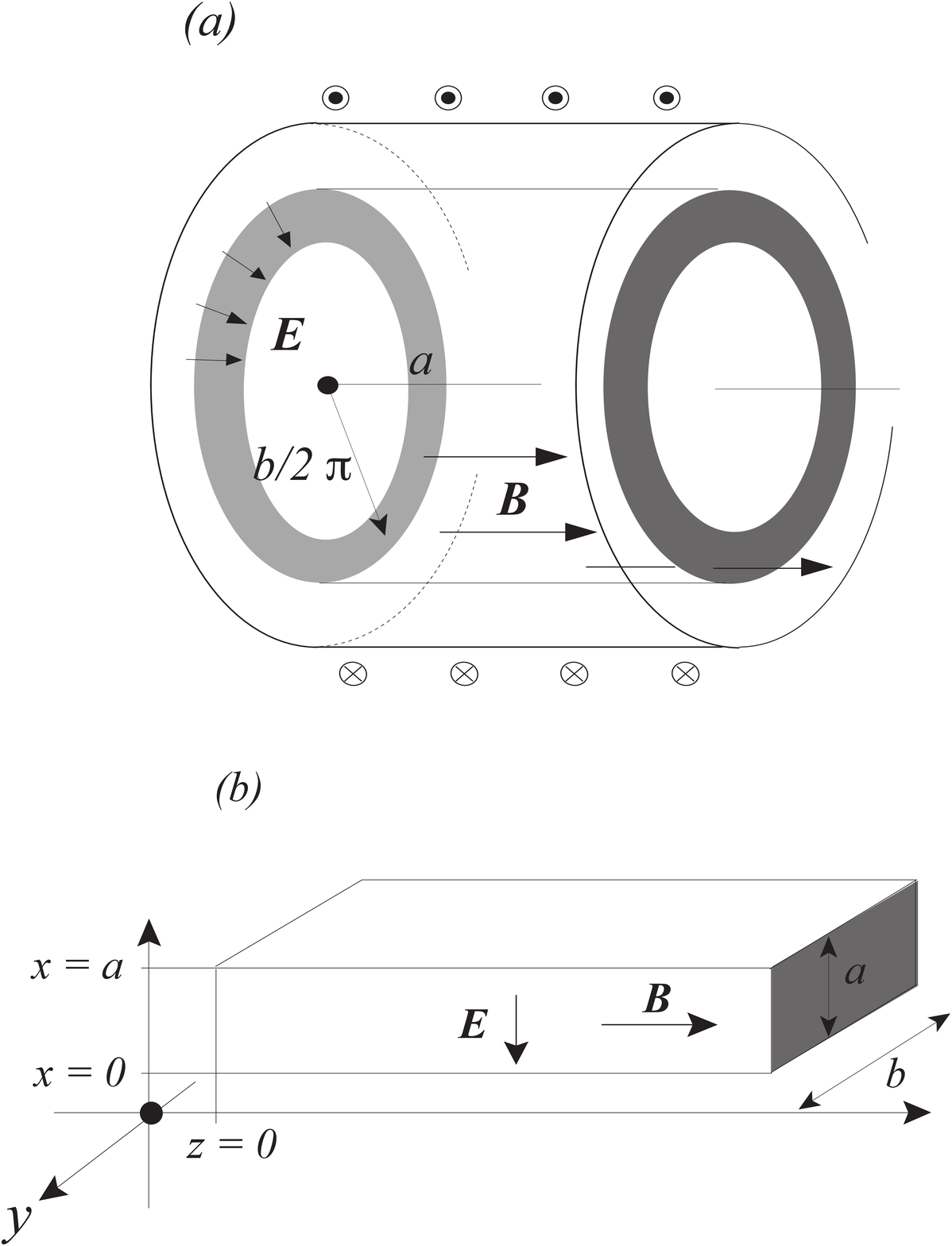}
\caption{Geometrical characteristics of the Cartesian plasma slab (b)
modeling the cylindrical plasma shell (a).}
\label{Fig:Fig4}
\end{center}
\end{figure}

Consider a cylindrical plasma shell uniformly magnetized along the $z$ axis.
In addition to the axial magnetic field $\mathbf{B}=B\mathbf{e}_{z}$ we consider a
radial electric field generated in a cylindrical shell of magnetic field
lines, with width $a$ and radius $b/2\pi $, depicted in grey on Fig.~\ref{Fig:Fig4}(a).
The radial electric field is generated in this cylindrical shell to sustain
a rotation around the $z$ axis for the purpose of thermonuclear confinement
or mass separation

\begin{figure}
\begin{center}
\includegraphics[width=8.6cm]{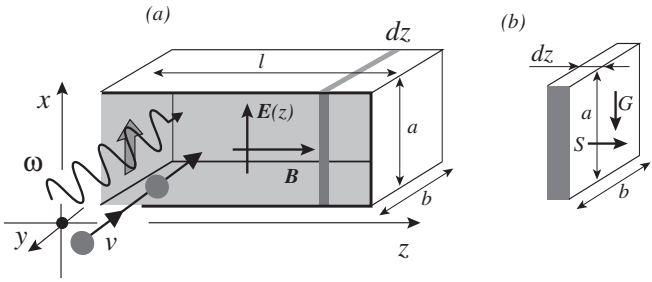}
\caption{(a) A plasma slab magnetized along $z$ and polarized along $x$
through wave/beam power absorption at $z=0$. (b) An infinitesimal slice $dz$
is fully charaterized by its transverse conductance $Gdz$ and longitudinal
resististance $Rdz$.}
\label{Fig:Fig5}
\end{center}
\end{figure}

In order to simplify the analysis, which can be also carried in cylindrical
coordinates, we will neglect curvature effects ($b>a$) and describe the grey
plasma zone of Fig.~\ref{Fig:Fig4}(a) as a slab plasma depicted in Fig.~\ref{Fig:Fig4}(b). This
transformation is just an unfolding of the cylindrical shell and displays
the advantage of simplifying the physical picture and results. Following
this unfolding, the Cartesian plasma slab considered in the following is
both magnetized along $z$, $\mathbf{B}=B\mathbf{e}_{z}$, and polarized along 
$x$, $\mathbf{E}=-E\mathbf{e}_{x}$.The magnetized plasma slab is of finite
size : (\textit{i}) $a$ along $x$, (\textit{ii}) $b$ along $y$ and (\textit{%
iii}) $l$ along $z$, as illustrated in Fig.~\ref{Fig:Fig5}(b).

The electric field is described by an electrostatic potential $V$ such that $%
\mathbf{E}$ = - $\left( \partial V/\partial x\right) \mathbf{e}_{x}$ - $\left( \partial V/\partial z\right) 
\mathbf{e}_{z}$ where $\partial V/\partial z<\partial V/\partial x=V/a$. The equivalent DC current
generator (wave or beam), located at $z=0$, sustains a current between $x=0$ and $x=a$. As a result of charges depletion at $x=0$ and charges accumulation at $%
x=a$, a voltage drop $V_{0}$ = $V\left( z=0\right) $ is sustained between
the magnetic surfaces $x=0$ and $x=a$. This voltage drop will decay away for 
$z>0$ because of the finite conductivities along $z$ and across $x$. These
finite conductivities will provide a fast dispersion of the charges along $z$
and a slow relaxation across $\mathbf{B}$ along $x$.

We assume (\textit{i}) that the amplitude of the wave is shaped such that
the wave equivalent current generator is driven from $x=0$ up to $x=a$ near $%
z=0$ and (\textit{ii}) that the density of the neutral beam is shaped such
that the beam equivalent voltage generator sets up a voltage drop between $x=0
$ and $x=a$ near $z=0$. In order to describe dissipative processes in the
slab $z>0$, we consider an infinitesimal slice of magnetized plasma : $dz$
along $z$, $a$ along $x$ and $b$ along $y$. This elementary slab, depicted
on Fig.~\ref{Fig:Fig5}(b), displays two properties: (\textit{i}) a large conductivity
along $dz$ and (\textit{ii}) a large resistivity along $x$. We assumed
cylindrical symmetry of the original problem which translates into
homogeneity along $y$ of the unfolded slab. In particular, as the wave and
beam travel in the $y$ direction, we assume homogeneous wave or beam power
deposition along $y$ near $z=0$, which means homogeneous current generation
and electric field generation along $y$.

\begin{figure}
\begin{center}
\includegraphics[width=8.6cm]{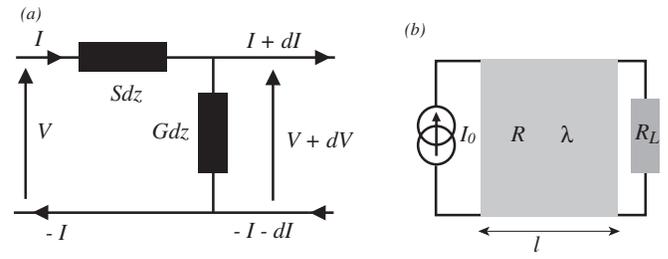}
\caption{(a) Equivalent circuit of a ($a,b,dz$) slice of the plasma. (b)
Equivalent model of power absorption and charge separation near $z=0$ and
charge distribution in the plasma slab terminated with loaded endplates at $%
z=l$. }
\label{Fig:Fig6}
\end{center}
\end{figure}

We describe the dissipative dynamics of the charges by the current $I\left(
z\right) $ which flow easily along $z$ and the small short circuited current
resulting from the small conductivity along $x$. In a slice $dz$ this short
circuiting of the initial charges separation is described by $dI/dz$. This
model allows to describe the volume charges relaxation and the steady state
large voltage drop generation across the magnetic field. To calculate the
small conductivity $Gdz$ along $x$ (across $B$) and the small resistivity $%
Sdz$ along $z$ (along $B$) we apply the classical formula describing the
resistance/conductance of the elementary parallelepiped depicted in Fig.~\ref{Fig:Fig5}(b), 
\begin{eqnarray}
Sdz &=&\frac{dz}{\eta _{\shortparallel }ba}\text{,}  \label{rr} \\
Gdz &=&\frac{\eta _{\perp }bdz}{a}\text{,}  \label{gg}
\end{eqnarray}
where we have introduced the classical conductivities $\eta _{\shortparallel
}$ and $\eta _{\perp }$ along and across the field lines in a magnetized
plasma~\cite{Helander2005,Rax2005,Gueroult2019b,Poulos2019,Rax2011,Trotabas2022}. Note that taking into account curvature effects would change the expression of $G$ but not $S$, with for the cylindrical shell illustrated on Fig.~\ref{Fig:Fig4}(a)
\begin{equation}
G=2\pi \eta _{\perp }/\ln \frac{1+(\pi a/b)}{1-(\pi a/b)},
\end{equation}
and we recover the previous expression if $a\ll b$. Then we apply Ohm's law to the transmission line like model illustrated in Fig.~\ref{Fig:Fig6}(a) to write the equations fulfilled by the
voltage $V$ across $x$ and the current $I$ along $z$ : 
\begin{eqnarray}
dV &=&-SIdz\text{,}  \label{tt1} \\
dI &=&-GVdz\text{.}  \label{tt2}
\end{eqnarray}
In order to obtain the various scalings and order of magnitude estimates of the final
results we use the classical formula for the longitudinal and transverse
conductivities used in Eqs.~(\ref{rr}, \ref{gg}).

Assuming first that the plasma is
not fully ionized and that collisions with neutrals at rest are the dominant
dissipative process : 
\begin{equation}
\eta _{\shortparallel }=\frac{n_{m}q^{2}}{m\nu _{m}}\text{, }\eta _{\perp }=%
\frac{n_{M}Q^{2}\nu _{M}}{M\omega _{c}^{2}}.
\end{equation}
Here $n$ is the density of free charges with mass $m$ (electrons) or $M$
(ions) and charges $q$ or $Q$, $\eta _{\shortparallel }$ is associated with
the electron population and $\eta _{\perp }$ with the ion one, and $n_{M}Q=n_{m}q$. The collision frequency $\nu $ can be either the collision frequency with neutrals in a cold
plasma or the turbulent decorrelation frequency in a turbulent plasma.

On the other hand, if the plasma is fully ionized, the conductivity along the field lines is
given by the Spitzer conductivity. It is independent of the density but
scales as $T^{-3/2}$ with the temperature, 
\begin{equation}
\eta _{\shortparallel }=\varepsilon _{0}\frac{\omega _{pe}^{2}}{\nu _{ei}}%
\text{.}  \label{ffspi}
\end{equation}
Across the field lines no relative velocity between electrons and ions is
observed in the $\mathbf{E}\times \mathbf{B}$ rest frame. This means that we
have to consider additional effect to find a dissipative channel. Among
these processes (\textit{i}) inertia, (\textit{ii}) viscosity and (\textit{%
iii}) inhomogeneity are usually put forward~\cite{Rozhansky2008,Rax2019,Kolmes2019}. We will consider here the
effect of inhomogeneity which displays the same scaling as viscosity~\cite{Rozhansky2008}.
In an inhomogeneous electric field, the expression of the electric drift
velocity $\mathbf{v}_{E\times B}$ is given by : 
\begin{equation}
\mathbf{v}_{E\times B}=\left( 1+\frac{\rho ^{2}}{4}\frac{d^{2}}{dx^{2}}%
\right) \frac{\mathbf{E}\times \mathbf{B}}{B^{2}}
\end{equation}
where $\rho $ is the Larmor radius. We will assume $d^{2}E/dx^{2}\sim
E/a^{2} $. This velocity is along $y$ and, because of the difference in
Larmor radius $\rho _{e}\ll \rho _{i}$, Coulomb collisions, at a rate $\nu
_{ie}$, provides a friction force $F$ between the electron and ion
populations. As a result the ion population experiences an $y$ directed force 
$F$ 
\begin{equation}
F=\nu _{ie}\frac{k_{B}T_{i}}{4\omega _{ci}^{2}}\frac{E}{a^{2}B}
\end{equation}
where $\omega _{ci}$ is the ion cyclotron frequency. This force $F$ along $%
y $ is the source of a $\mathbf{F}\times \mathbf{B}/QB^{2}$ drift along $x$
and this drift gives the equivalent conductivity $\eta _{\perp }$ associated
with inhomogeneity : 
\begin{equation}
\eta _{\perp }=n_{i}\frac{\nu _{ie}}{\omega _{ci}}\frac{\rho _{i}^{2}}{a^{2}}%
\frac{Q}{4B}=\frac{\varepsilon _{0}}{4}\nu _{ie}\frac{\omega _{pi}^{2}}{%
\omega _{ci}^{2}}\frac{\rho _{i}^{2}}{a^{2}}.  \label{fffrt}
\end{equation}

The strong scaling with respect to the magnetic field $\rho _{i}^{2}/\omega
_{ci}^{2}\sim B^{-4}$ is to be noted. The effect of viscosity displays the
same scaling and we will consider Eq.~(\ref{fffrt}) as the approximate
perpendicular conductivity of a fully ionized plasma~\cite{Rozhansky2008}. In the
following, to evaluate the power dissipation with Eq.~(\ref
{ffspi}, \ref{fffrt}), we will use the following estimate for a fully ionized hydrogen
plasma 
\begin{equation}
\nu _{ei}=\ln \Lambda \left[ \frac{mc^{2}}{3k_{B}T}\right] ^{\frac{3}{2}}%
\frac{r_{e}}{c}\omega _{pe}^{2}\sim \left[ \frac{mc^{2}}{3k_{B}T}\right] ^{%
\frac{3}{2}}\left[ \frac{\omega _{pe}}{10^{11}\text{Rd/s}}\right] ^{2}
\end{equation}
where $r_{e}=2.8\times 10^{-15}$ m is the classical electron radius, $mc^{2}$
$=511$ KeV the electron rest energy and $c=2.9\times 10^{8}$ m/s the
velocity of light. The ion-electron collision frequency is given by $\nu
_{ie}$ = $m\nu _{ei}/M$.

\section{Attenuation length and plasma resistance}
\label{Sec:SecVI}

In order to analyze Eqs.~(\ref{tt1}, \ref{tt2}), it turns out to be more
convenient to introduce what we will call the \textit{plasma slab resistance}
$R$ defined as 
\begin{equation}
Rb=\frac{1}{\sqrt{\eta _{\perp }\eta _{\shortparallel }}}\text{,}
\label{Rpla}
\end{equation}
and the \textit{attenuation length} $\lambda $ defined as
\begin{equation}
\frac{\lambda }{a}=\sqrt{\frac{\eta _{\shortparallel }}{\eta _{\perp }}}%
\text{.}  \label{attl}
\end{equation}
These two global characteristics, $R$ and $\lambda $, capture all the
electrical properties of the plasma slab needed to describe the charge
relaxation for $z>0$ of the $z=0$ wave or beam driven perpendicular current.

For a fully ionized plasma the transverse conductivity is a second order
effect described by Eq.~(\ref{fffrt}) and the plasma resistance and
attenuation length are given by
\begin{eqnarray}
\frac{\lambda }{a} &=&\frac{2}{\sqrt{\nu _{ei}\nu _{ie}}}\frac{\omega
_{pe}\omega _{ci}}{\omega _{pi}}\frac{a}{\rho _{i}}\sim \frac{\omega _{ci}}{%
\nu _{ie}}\frac{a}{\rho _{i}}  \label{Rpla2} \\
\frac{1}{Rb} &=&\frac{\varepsilon _{0}}{2}\frac{\omega _{pe}\omega _{pi}}{%
\omega _{ci}}\sqrt{\frac{\nu _{ie}}{\nu _{ei}}}\frac{\rho _{i}}{a}\sim
\varepsilon _{0}\frac{\omega _{pe}^{2}}{\omega _{ce}}\frac{\rho _{i}}{a}
\label{Rpla3}
\end{eqnarray}
The attenuation length $\lambda $ is thus far larger than the size of the
device for a fully ionized plasma of the thermonuclear type. Note also that while the definition of the attenuation length $\lambda $, Eq.~(\ref{attl}%
), already appears in the literature in the few studies addressing the
issue of field penetration from the edge~\cite{Gueroult2019b,Poulos2019,Trotabas2022}, the definition of 
\begin{equation}
R = \frac{\omega _{ce}a}{b\varepsilon _{0}\rho _{i}\omega _{pe}^{2}}\nonumber
\end{equation}
for a fully ionized plasma, Eq.~(\ref{Rpla3}), does not seem to have attracted some
previous specific attention despite its importance to understand DC voltage
distribution in a fully ionized magnetized plasma. 

With these definitions
Eqs.~(\ref{tt1}, \ref{tt2}) become simply 
\begin{eqnarray}
\lambda \frac{dV}{dz} &=&-RI\text{,}  \label{ttt1} \\
\lambda \frac{dI}{dz} &=&-\frac{V}{R}\text{.}  \label{ttt2}
\end{eqnarray}
We further define the new variables $s=z/\lambda $ and $\left( u,v\right) $ such
that 
\begin{equation}
\left( 
\begin{array}{l}
u \\ 
v
\end{array}
\right) =\left( 
\begin{array}{l}
\frac{V}{\sqrt{R}}+\sqrt{R}I \\ 
\frac{V}{\sqrt{R}}-\sqrt{R}I
\end{array}
\right) \text{,}
\end{equation}
so that 
\begin{equation}
\frac{d}{ds}\left( 
\begin{array}{l}
u \\ 
v
\end{array}
\right) =\left( 
\begin{array}{l}
-u \\ 
+v
\end{array}
\right) \text{.}  \label{dec1}
\end{equation}
The solutions of Eq.~(\ref{dec1}) which are simply a \textit{forward decay} $%
u=u_{0}\exp -s$ and a \textit{backward decay} $v=v_{0}\exp s$.

Note for completeness that Eq.~(\ref{dec1}) was derived by assuming that the plasma is homogeneous. A
simple model taking into account the $z$ variation of $\lambda \left(
z\right) $ and $R\left( z\right) $ can be studied in a way similar to the
analysis of the previous homogeneous model but by considering this time the change of
variable 
\begin{equation}
s\left( z\right) =\int_{0}^{z}du/\lambda \left( u\right) \text{.}
\label{ttuiop2}
\end{equation}
With this change of variables Eq.~(\ref{dec1}) becomes 
\begin{equation}
\frac{d}{ds}\left( 
\begin{array}{l}
u \\ 
v
\end{array}
\right) =\left( 
\begin{array}{l}
-u \\ 
+v
\end{array}
\right) -\left( d\ln \sqrt{R}/ds\right) \left( 
\begin{array}{l}
v \\ 
u
\end{array}
\right) \text{.}  \label{inh6}
\end{equation}
and the forward and backward solution are coupled by the inhomogeneities.
This inhomogeneities $\lambda \left( z\right) $ and $R\left( z\right) $ play
the role of an additional dissipative term, for example, when the magnetic
field lines are diverging. Although interesting generalizations, the tapering effect of inhomogeneous plasma and
magnetic field properties will not be considered here, and we will restrict the
analysis to the solutions of Eq.~(\ref{dec1}).

The general solution of Eqs.~(\ref{ttt1}, \ref{ttt2}) is a linear combination
of the forward and backward solutions $\exp +z/\lambda $ and $\exp
-z/\lambda $. In the following we consider the general solution
\begin{eqnarray}
I\left( z\right) &=&I_{-}\exp \left( -\frac{z}{\lambda }\right) +I_{+}\exp
\left( +\frac{z}{\lambda }\right)  \label{tele1} \\
V\left( z\right) &=&RI_{-}\exp \left( -\frac{z}{\lambda }\right) -RI_{+}\exp
\left( +\frac{z}{\lambda }\right)  \label{tele2}
\end{eqnarray}
where the amplitudes $I_{\pm }$ are given by the two boundary conditions (%
\textit{i}) at $z=0$ with the wave or beam driven generators, and (\textit{%
ii}) at $z=l$ with a load $R_{L}$ describing how
we choose to terminate the field lines and the plasma. This is illustrated in Fig.~\ref{Fig:Fig6}(b). The $\exp +z/\lambda $
solution is associated with the reflection on the load at $z=l$ when there
is an impedance mismatch of this load $R_{L}$ with the plasma resistance $R$.

The boundary condition at $z=0$ depends on whether wave or neutral
beam is considered. For the wave case, as the
effect of the wave is to move already existing charges, we consider an
equivalent perfect current generator $\left. I_{0}\right| _{RF}$ localized
at $z=0$. For the neutral beam case, as the beam brings and
separates charges with opposite signs, we consider an equivalent perfect
voltage generator $\left. V_{0}\right| _{NB}$ localized at $z=0$. We call $%
I_{0}=I\left( z=0\right) $ the current of the generator equivalent to the
wave, and $V_{0}=V\left( z=0\right) $ the voltage drop in the beam
active region near $z=0$. These current and voltage generators can be respectively related to the injected RF power and beam momentum as follows.  

Writing $\mathcal{P}_{RF}\left[ \text{W}\right] $ the total power absorbed by the plasma from the wave at $z=0$ where the wave power
deposition is localized, one gets
\begin{equation}
P_{RF}\left[ \text{W/m}^{3}\right] =\frac{\mathcal{P}%
_{RF}\delta \left( z\right)}{ab}
\end{equation}
where $\delta \left( z\right) $ is the
Dirac distribution. Then from Eq.~(\ref{current}) we can define the
equivalent current generator $\left. I_{0}\right| _{RF}$ associated with the wave
drive at $z=0$ through the relation $J_{\perp }$ = $%
\left. I_{0}\right| _{RF}\delta \left( z\right) /b$, so that
\begin{equation}
\left. I_{0}\right| _{RF}=\frac{k_{\perp }}{\omega }\frac{1}{Ba}\mathcal{P}%
_{RF}.  \label{II2}
\end{equation}
Similarly, we can define from Eq.~(\ref{ebeam 2}) the equivalent voltage
generator$\left. V_{0}\right| _{NB}$ = $E_{NB}a$ associated with the beam
drive at $z=0$%
\begin{equation}
\left. V_{0}\right| _{NB}=aB\nu \tau \frac{N_{B}}{N_{p}}v.  \label{III4}
\end{equation}
For the wave case the power of the wave equivalent generators is $\left.
I_{0}\right| _{RF}V_{0}$. Under optimal conditions such as discussed in section~\ref{Sec:SecII},
energy conservation implies that the input RF power is equal to the dissipated DC
power : $\left. I_{0}\right| _{RF}V_{0}$ = $\mathcal{P}_{RF}$. Eliminating $%
\mathcal{P}_{RF}$ between this last relation and Eq.~(\ref{II2}) we recover
Eq.~(\ref{estar}) as expected.

Because of dissipation the current $\left. I_{0}\right| _{RF}$ and voltage $%
\left. V_{0}\right| _{NB}$ are progressively shunted by the plasma, away
from $z=0$, as a result of the high conductivity along $z$ and the weak
conductivity along $x$. This decrease is described by the solution Eqs.~(\ref
{tele1}, \ref{tele2}) under the appropriate boundary conditions $I\left(
z=0\right) =I_{0}$ or $V\left( z=0\right) =V_{0}$ given by Eqs.~(\ref{II2}, 
\ref{III4}) and $V\left( z=l\right) $ = $R_{L}I\left( z=l\right) $ at the
end of the field lines for a plasma column of length $l$.

\section{Power dissipation in a loaded plasma slab}
\label{Sec:SecVII}

\subsection{Power requirement}

We consider Eqs.~(\ref{tele1}, \ref{tele2}) with the wave or beam driven
generator Eq.~(\ref{II2}) or Eq.~(\ref{III4}) at $z=0$, and with the plasma being
terminated at $z=l$ by a resistive load $R_{L}$ as illustrated on Fig.~\ref{Fig:Fig6}(b). These boundary conditions can be written as
\begin{equation}
I_{-}+I_{+} =I_{0} \label{tttui1} 
\end{equation}
and
\begin{multline}
R\left( I_{-}\exp -\frac{l}{\lambda }-I_{+}\exp +\frac{l}{\lambda }\right)
\\=R_{L}\left( I_{-}\exp -\frac{l}{\lambda }+I_{+}\exp +\frac{l}{\lambda }%
\right).\label{ttui2}
\end{multline}
%\begin{eqnarray}
%I_{-}+I_{+} &=&I_{0}  \label{tttui1} \\
%R\left( I_{-}\exp -\frac{l}{\lambda }-I_{+}\exp +\frac{l}{\lambda }\right)
%&=&R_{L}\left( I_{-}\exp -\frac{l}{\lambda }+I_{+}\exp +\frac{l}{\lambda }%
%\right)  \label{ttui2}
%\end{eqnarray}
After some elementary algebra, we solve Eqs.~(\ref{tttui1}, \ref{ttui2}) for
the amplitudes $I_{\pm }$ and express $V\left( z=0\right) $ as a function of 
$I\left( z=0\right) $ through the definition of $R_{e}$: $V_{0}=R_{e}I_{0}$. This resistance $R_{e}$ is the equivalent resistance of the plasma slab as seen from $z=0$, and writes 
\begin{equation}
\frac{R_{e}}{R}=\frac{R_{L}+R\tanh l/\lambda }{R+R_{L}\tanh l/\lambda }\text{%
.}  \label{equi}
\end{equation}
For the wave case, Eq.~(\ref{II2}) relates the current $\left. I_{0}\right|
_{RF}$ to the RF power $\mathcal{P}_{RF}$. This power is used to sustain the
steady state current and voltage pattern in the plasma slab ($a,b,l$)
against relaxation. The maximum voltage drop in the wave active region $z=0$
is thus 
\begin{equation}
\left. V_{0}\right| _{RF}=R_{e}\frac{k_{\perp }}{a\omega B}\mathcal{P}%
_{RF}\leq \frac{R}{\tanh l/\lambda }\frac{k_{\perp }}{a\omega B}\mathcal{P}%
_{RF}  \label{powx1}
\end{equation}
where the right hand side of the inequality, $R_{e}=R/\tanh l/\lambda $, is
associated withe optimal choice for the load at $z=l$, that is $R_{L} \rightarrow +\infty $. As $\tanh l/\lambda $ increases from zero up to one when $l$ increases,
a shorter plasma column displays a larger voltage drop for the same power because
the charges are more concentrated on the field lines, in the limit that $l<\lambda $. With
the expansion: 
\begin{equation}
\left. R_{e}\right| _{R_{L}\rightarrow+\infty }=\frac{R}{\tanh l/\lambda }\approx 
\frac{\lambda R}{l}=\frac{a}{bl\eta _{\perp }}\text{,}
\end{equation}
the plasma slab behaves as an isotropic conductor with conductivity $\eta
_{\perp }$ and Eq.~(\ref{powx1}) becomes : 
\begin{equation}
\left. V_{0}\right| _{RF}\approx \frac{k_{\perp }}{bl\eta _{\perp }\omega B}%
\mathcal{P}_{RF}  \label{powx4}
\end{equation}
Dissipation across the field lines is ultimately responsible for the limit
described by Eq.~(\ref{powx4}). For such a favorable limit, even if $\eta
_{\perp }\rightarrow 0$ or $\mathcal{P}_{RF}\rightarrow +\infty $ the
optimum voltage $V_{0}$ is limited by the relation Eq.~(\ref{estar}) which
is a constraint imposed by the wave-particle resonance if we want to
optimize the generation process and avoid to waste power into Landau and
cyclotron heating.

Using Eq.~(\ref{fffrt}) the power requirement $\mathcal{P}$ $%
\sim $ $bl\eta _{\perp }V_{0}^{2}/a$ for a given voltage drop and a given
fully ionized plasma under optimal conditions is
%\begin{widetext}
%\begin{equation}
%\left[ \frac{\mathcal{P}}{1~\text{W}}\right] \sim \left[ \frac{V_{0}}{%
%10^{10}~\text{V}}\right] ^{2}\left[ \frac{T}{1~\text{eV}}\right] ^{-\frac{3%
%}{2}}\left[ \frac{n_{e}}{1~\text{cm}^{-3}}\right] \left[ \frac{l}{1~\text{m%
%}}\right]\ln \Lambda \text{ }\frac{M}{m}\frac{\rho _{i}^{2}}{a^{2}}\frac{b}{a%
%}\frac{\omega _{pe}^{2}}{\omega _{ce}^{2}}\text{.} \label{pow678}
%\end{equation}
%\end{widetext}
\begin{widetext}
\begin{equation}
\left[ \frac{\mathcal{P}}{\text{W}}\right] \sim \left[ \frac{V_{0}}{%
~\text{MV}}\right] ^{2}\left[ \frac{\omega_{{pe}}}{%
10^{11}~\text{rad.s}^{-1}}\right]^{2}\left[ \frac{l}{\text{m%
}}\right]\left[\frac{b}{a}\right]\left[ \frac{k_{B}T}{mc^{2}}\right]^{-\frac{3%
}{2}}\left[\frac{\rho_{i}}{a}\right]^{2}\left[\frac{\omega_{pe}}{\omega_{ce}}\right]^{2}, \label{pow678}
\end{equation}
\end{widetext}
where we assumed $\ln \Lambda=10$. This result suggests that megavolt voltage drops are accessible for rather low driving power in thermonuclear hydrogen plasmas where typically $b\sim a$, $\omega
_{pe}\sim \omega _{ce}$ and $a\geq 10\rho _{i}$.

\begin{figure}
\begin{center}
\includegraphics[width=8.6cm]{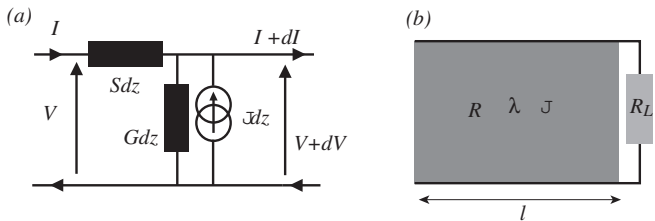}
\caption{ (a) Equivalent circuit
of a $dz$ slice $\left( a,b,dz\right) $ of the plasma. (b) Equivalent model
of wave absorption and charge separation and charge dissipation in the
plasma slab $\left( a,b,l\right) $ terminated with loaded endplates at $z=l$. }
\label{Fig:Fig7}
\end{center}
\end{figure}

Up to now we have only considered a current source (equivalent to the wave
or the beam) localized near $z=0$. For wave drive this is true if the resonant
particles are chosen with a zero parallel velocity, and/or if
the plasma column is very long, and/or if the quasilinear wave diffusion
from $x=0$ to $x=a$ is fast enough compared to the other processes. This
issue of the radial current deposition by a wave must be addressed within
the framework of a collisional/quasilinear kinetic model. Similarly the
issue of the neutral beam current deposition is to be addressed within a
kinetic model. Rather than going this route we consider here for completeness the previous fluid model but the complementary and more general problem of a broad current
deposition profile. Specifically, the wave or beam current deposition is assumed to be broadly distributed all along the field lines, $0<x<l$, and described by an infinitesimal current source, $\mathcal{I}dz$ = $\left( I_{0}/l\right) dz$, in each infinitesimal section $dz$ along $z$. We consider the
equivalent circuit associated with an infinitesimal section $dz$ as illustrated in Fig.~\ref{Fig:Fig7}(a). The electrical properties of a slice $\left( a,b,dz\right) $ then take into account a $\mathcal{I}dz$ current source. 

The transmission line equations describing the slab $\left( a,b,l\right)$ with load $R_{L}$ at $z=l$
as illustrated in Fig.~\ref{Fig:Fig7}(b) are 
\begin{align}
\lambda \frac{dV}{dz} =&-RI\text{, }  \label{esd1} \\
\lambda \frac{dI}{dz} =&-\frac{V}{R}+\lambda \mathcal{I}\text{.}
\label{esd2}
\end{align}
Note that Eqs.~(\ref{esd1}, \ref{esd2}) will still hold true if considering
plasma conductivities and power deposition profiles that are inhomogeneous
along $z$. With the boundaries conditions $I\left( z=0\right) $ = $0$ and $R_{L}I\left(
z=l\right) =V\left( z=0\right) $, the solutions are given by 
\begin{align}
I\left( z\right)  =&\mathcal{I}\lambda \frac{R\sinh \left( z/\lambda
\right) }{R\cosh \left( l/\lambda \right) +R_{L}\sinh \left( l/\lambda
\right) }\text{,} \\
V\left( z\right)  =&R\mathcal{I}\lambda \left[ 1-\frac{R\cosh \left(
z/\lambda \right) }{R\cosh \left( l/\lambda \right) +R_{L}\sinh \left(
l/\lambda \right) }\right] \text{.}  \label{vovo}
\end{align}
With these solutions we can now define two equivalent resistances. The first one
is simply the ratio of the voltage $V_{0}$ = $V\left( z=0\right) $ to the
total wave or beam driven current $I_{0}$ = $\int_{0}^{l}\mathcal{I}dz$,%
\begin{equation}
\frac{V_{0}}{I_{0}}  =  R\frac{\lambda }{l}\left[ 1-\frac{R}{R\cosh \left(
l/\lambda \right) +R_{L}\sinh \left( l/\lambda \right) }\right] \underset{R_{L}\rightarrow +\infty }{\approx} R\frac{\lambda }{l}.
\end{equation}
The second resistance is more instructive and is associated with the
integrated global power balance 
\begin{equation}
R_{e}^{\prime }=\frac{\int_{0}^{l}V\left( z\right) \mathcal{I}dz}{\left(
\int_{0}^{l}\mathcal{I}dz\right) ^{2}}\text{.}
\end{equation}
Indeed, similarly to what was discussed for the localised source, this is this resistance $R_{e}^{\prime }$ which now determines the power balance of the wave or beam driven rotation
process for a broad power deposition profile. 
Using Eq.~(\ref{vovo}) this resistance rewrites
\begin{equation}
R_{e}^{\prime }=R\frac{\lambda }{l}\left[ 1-\frac{\lambda }{l}\frac{R\sinh
\left( l/\lambda \right) }{R\cosh \left( l/\lambda \right) +R_{L}\sinh
\left( l/\lambda \right) }\right].
\end{equation}
Interestingly, we find that
\begin{equation}
R_{e}^{\prime } \underset{R_{L}\rightarrow +\infty }{\approx} R\frac{\lambda }{l},
\end{equation}
so that the same result is obtained for distributed and localized drives under optimal condition $R_{L}\rightarrow+\infty $. In other words, the power requirement is rather
insensitive to the current deposition profile along field lines $0\leq x\leq l$ when $R_{L}\rightarrow +\infty $ or $%
l<\lambda $. 

\subsection{Voltage shaping}

Besides the power requirement, the model developed here can also be used to study the voltage shaping issue. Indeed, while a careful shaping of the radial power deposition
profile can be used to control the radial structure of the electric field,
its axial structure is determined by the plasma properties $\lambda $,
and strategies to control this axial distribution are to be identified. An issue here is that while the assumption $\eta _{\shortparallel }=\eta _{\text{%
Spitzer}}$ is confirmed by experiments in fully ionized plasmas, there exists no large experimental
data basis for $\eta _{\perp }$ in fully ionized, magnetized, (supersonic)
rotating plasmas. As a result, we can not accurately calculate the attenuation length $\lambda $ and the resistance $R_{e}$ in a fully ionized plasma column of length $l$. We can however, as we will do now, identify trends. 

Consider first the limit $\lambda >l$. In this limit the plasma column is not highly dissipative
and the power needed to sustain a large radial electric field is small if $%
R_{L}$ is large. The large voltage drop is however to be handled at the left and
right edge of the column with concentric circular end plates, and the issue
of the management of high voltage between conductors must then to be solved. Consider now the opposite limit $\lambda <l$. In this limit the plasma column is rather dissipative and
the power needed to sustain a large radial electric field will be large.
On the other hand the insulation of the endplates terminating the field
lines will not be a problem. The former situation, that is limited dissipation $\lambda >l$, is the one we will focus on in the remaining of this section.

\begin{figure}
\begin{center}
\includegraphics[width=8.6cm]{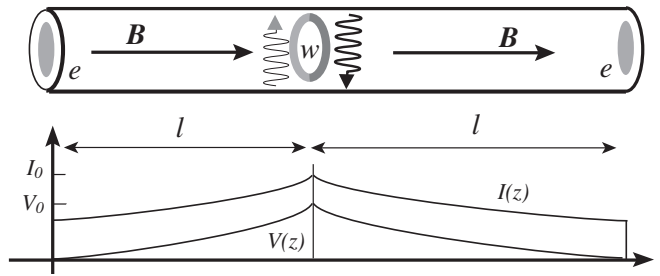}
\caption{A magnetized plasma column with two ergodized zone ($e$) and a
central wave/beam driven zone ($w$).}
\label{Fig:Fig8}
\end{center}
\end{figure}

Consider a plasma column of length $l$ as illustrated in Fig.~\ref{Fig:Fig8}. The wave driven
current generator $I_{0}=\mathcal{P}_{RF}k_{\perp }/a\omega B$ is assumed to be localized
around $z=0$ ($w$), and the transverse conductivity $\eta
_{\perp }$ is assumed to become very large near $z=\pm l$. This end zone $(e)$  in Fig.~\ref{Fig:Fig8} can be considered as a short circuit such that $R_{L}=0$. With these two boundaries conditions, $V\left(
z=l\right) =0$ and $I\left( z=0\right) =I_{0}$, and focusing on the region $z>0$, the solutions Eqs.~(\ref
{tele1}, \ref{tele2}) give
\begin{align}
I\left( z\right) & =I_{0}\cosh \frac{l-z}{\lambda }\left(\cosh \frac{l}{%
\lambda }\right)^{-1},\\
V\left( z\right) & = RI_{0}\sinh \frac{l-z}{\lambda }\left(%
\cosh \frac{l}{\lambda }\right)^{-1}.  \label{VVV}
\end{align}
Symmetrical solutions are expected for $z<0$, as illustrated
in Fig.~\ref{Fig:Fig8}. Note also that we should take $2I_{0}$ as the wave driven current flows both on the left and right sides of the central region $(w)$. 

Although the important problem of how to implement the condition $R_{L}=0\ $at $z=\pm l$ is left for a future study, we briefly discuss here local ergodization of the magnetic field lines. The required magnetic modulations can be achieved with external coils producing radial and azimuthal components of the magnetic field. The magnetic field lines then display the property of being an Hamiltonian system where the time is replaced by the $z$ coordinate, so that if the
local modulations have several resonances and enter the regime where the
Chirikov criterion is fulfilled. The field lines, which are basically
the wire along which the free charges flow, will then explore the full radial
extent of the zone depicted in grey (e) on Fig.~\ref{Fig:Fig8}, which will provide an
almost perfect short circuit between $x=0$ and $x=a$ in the slab model.
Ergodization of magnetic field lines is common in plasma physics and
particularly in tokamak plasma where the principle of magnetic island
overlapping has been put forward and tested successful with the concept of 
\textit{ergodic divertor}. Yet, the use of this strategy for the problem at hand raises two
problems. First, the short circuit at $z=l$ implies that the power
needed to sustain the radial electric field to be very large. From Eq.~(\ref{VVV}), the power sustaining the generation and confinement of the electric field is 
\begin{equation}
I_{0}V_{0}\approx \frac{RI_{0}^{2}l}{\lambda} =\frac{I_{0}^{2}l}{ab\eta _{\shortparallel }}
\label{POW}
\end{equation}
The plasma slab thus behaves as an isotropic conductor with conductivity $\eta
_{\shortparallel }$. Second, it is not clear that an ergodic zone near the endplates will really protect them from damages as the short circuit will be the source of an
intense Joule heating. 

Beyond ergodization, alternative strategies to minimize the risk of high voltage damages at the edges of the plasma and to lower the power requirement will have to be established on the
specific material and power constraints of each configuration. Eq.~(\ref{equi}) provides the basis for such analysis. For very large electric fields, and if we let some part of the voltage drop
reach the end plates, a preferential combination of electrodes could possibly be used
to set up a classical energy recovery system outside the plasma. This part
of tolerable voltage will again have to be analyzed with respect to the electrodes
properties. Finally, we note that the occurrence of inhomogeneity described by Eq.~(\ref
{inh6}), such as the divergence of magnetic field lines, can in principle be used to
shape the axial voltage profile and reduce the electric field on the conducting plates. The examination of these possibilities is left for future studies.

\section{Discussion and conclusion}
\label{Sec:SecVIII}

In this first study on wave and beam large electric field generation and
control in the core of a magnetized plasmas, we have derived and solved the
equation for the axial variation of the voltage drop. We identified $R$ and $%
\lambda $ as the control parameters of the problem. We then used these results to address the
issue of the power balance, and of field shaping in the asymptotic regime $l<\lambda $.

To summarize our findings:
\begin{itemize}
\item[(\textit{i})] We have identified, proposed and
analyzed two mechanisms for large DC electric field generation inside a
magnetized plasma: waves and neutral beams, which are control tools that are already routinely
used on modern tokamaks at power levels of the order of tens of Megawatts~\cite{Rax2011}. The relations Eq.~(\ref{estar}) and Eq.~(\ref{max5}) provide upper
bounds for the electric field theoretically achievable with these wave and beam schemes. These
upper bound are in the GV/m range, which authorizes to consider tens of MV/m
electric field generation in magnetized plasmas. 
\item[(\textit{ii})]We have set
up a model of the plasma stationary response to wave and beam power
absorption. This model predicts both the electric field penetration from the
edge in the classical scheme Fig.~\ref{Fig:Fig1}(a), and the electric field escape from
the core central part of a column in the wave or beam driven scheme Fig.~\ref{Fig:Fig1}(b) and Fig.~\ref{Fig:Fig1}(c).
\item[(\textit{iii})] We have derived the voltage drop equation for an
axially inhomogeneous plasma Eq.~(\ref{inh6}). 
\item[(\textit{iv})] We have
identified the three fundamental characteristics of a plasma slab: $R$, Eq.~(\ref{Rpla}), and $\lambda $, Eq.~(\ref{attl}), and then calculated the input
impedance of the plasma slab $R_{e}$, Eq.~(\ref{equi}). 
\item[(\textit{v})] We derived in Eq.~(\ref{pow678}) the
minimal power required to sustain a given voltage drop $%
\mathcal{P}a\sim $ $bl\eta _{\perp }V_{0}^{2}$, and showed that
MV/m fields are within the power range of existing
wave and beam control devices in large tokamak. 
\end{itemize}

To extend this set of new
results, other schemes to localize the voltage drop inside the plasma
column, far from the edge, can be explored on the basis of Eq.~(\ref{inh6})
which is to be completed by appropriate loading or biasing conditions at $%
s=\int_{0}^{\pm l}dz/\lambda \left( z\right) $.

\section*{Acknowledgments}

The authors would like to thank Dr. I. E. Ochs, E. J. Kolmes, T. Rubin, and
M. E. Mlodik for constructive discussions. This work was supported by ARPA-E
Grant No. DE-AR001554. JMR acknowledge Princeton University and the
Andlinger Center for Energy + the Environment for the ACEE fellowship which
made this work possible.

\section*{References}
%\bibliography{RefsJMR}
%merlin.mbs aipnum4-1.bst 2010-07-25 4.21a (PWD, AO, DPC) hacked
%Control: key (0)
%Control: author (8) initials jnrlst
%Control: editor formatted (1) identically to author
%Control: production of article title (-1) disabled
%Control: page (0) single
%Control: year (1) truncated
%Control: production of eprint (0) enabled
%

\end{document}